\documentclass[sigconf]{acmart}
\AtBeginDocument{%
  }

\acmISBN{978-1-4503-XXXX-X/20XX/XX}

\usepackage{caption}
\usepackage{multirow}
\usepackage{subcaption}
\usepackage{graphicx}
\usepackage{amsmath}
\usepackage{amsthm}
\usepackage{booktabs}
\usepackage{pifont}
\usepackage{color}
\usepackage{dblfloatfix}
\usepackage{tabularx}
\usepackage{algorithm}
\usepackage{algorithmic}
\usepackage{booktabs}
\usepackage{hyperref}

\newtheorem{theorem}{Theorem}[section]
\newtheorem{lemma}[theorem]{Lemma}

\DeclareRobustCommand{\cmark}{\text{\ding{51}}}
\DeclareRobustCommand{\xmark}{\text{\ding{55}}}

\begin{document}

\title{A-Tag: Cryptographic File Tagging with Enforced Live Human Presence}

\author{Mohammed Gharib}
\email{gharib@aermob.com}
\orcid{0000-0001-9611-1693}
\affiliation{%
  \institution{Aerendir Mobile Inc.}
  \city{Mountain View}
  \state{CA}
  \country{USA}
}

\author{Sam Burns}
\email{sam@aermob.com}
\orcid{}
\affiliation{%
  \institution{Aerendir Mobile Inc.}
  \city{Mountain View}
  \state{CA}
  \country{USA}
}

\author{Martin Zizi}
\email{martin@aermob.com}
\orcid{}
\affiliation{%
  \institution{Aerendir Mobile Inc.}
  \city{Mountain View}
  \state{CA}
  \country{USA}
}
\renewcommand{\shortauthors}{Gharib et al.}

\begin{abstract}
The increasing autonomy of software agents and the proliferation of AI-generated content have weakened the reliability of digital actions as evidence of human involvement. Existing mechanisms, such as digital signatures and authentication protocols, verify identity and integrity, but do not establish whether a live human was present at the time an action was performed. In this paper, we introduce \textit{A-Tag}, a system for \emph{verifiable human-in-the-loop tagging} of digital artifacts. A-Tag cryptographically binds a file to a verified instance of live human presence at the moment of tagging. The resulting tag serves as a persistent and transferable proof that the action was triggered by a live human, without making claims about the origin, identity of the actor, or semantic content of the file.
A-Tag is a modular framework combining liveness verification based on micro-motion signals shaped by human motor control, a secure cryptographic binding between the verified event and the artifact, and a publicly verifiable logging mechanism enabling independent validation. We formalize security properties of human presence signals, define an adversarial model capturing automated and relay attacks, and analyze the system’s guarantees and limitations.
We implement A-Tag on commodity devices and evaluate it under realistic conditions. Our evaluation shows that A-Tag provides cryptographically sound artifact binding with deterministic verification, while enabling robust recovery under real-world transformations, achieving over 99\% success for image-based artifacts, and maintaining sub-second verification and practical tagging latency.
\end{abstract}

\maketitle

\section{Introduction}

Modern digital systems are fundamentally reshaping how digital actions are produced, delegated, and consumed. They routinely generate, modify, and distribute digital artifacts—such as documents, messages, and decisions—without continuous human involvement or any observable signal of human presence. While this shift enables scalability and efficiency, it also exposes a critical gap in security-sensitive workflows: the absence of assurance that a real, live human was actively interacting with the device when an action was performed.
Existing security mechanisms, including authentication protocols and digital signatures, provide strong guarantees about identity, integrity, and non-repudiation, but they do not establish whether a human was physically present and actively interacting with the device at the time of an action. These mechanisms treat an action as valid if it is authorized, regardless of whether it was initiated by a human or executed autonomously. Consequently, credentials can be delegated to software agents, misused by malware, or exploited at scale by adversaries, allowing actions to appear legitimate without real-time human involvement.

For example, consider a system in which documents are automatically generated and signed using a valid device-resident credential. Although each document carries a correct digital signature, the signing operation can be invoked programmatically—by background services, malware, or automated pipelines—without any real-time human involvement. Thus, a valid signature alone cannot establish that a human was present when the signing operation was triggered.
This distinction is increasingly consequential in settings involving automated content generation, bot-driven amplification, and delegated decision-making pipelines. In such environments, the ability to separate human-triggered actions from purely automated ones becomes critical. Importantly, the goal is not to restrict benign automation, but to enable systems in which human presence can be selectively enforced when accountability is required.

In this paper, we introduce \textit{A-Tag}, a system for \emph{verifiable human-in-the-loop tagging} of digital artifacts. A-Tag computes a cryptographic digest of the target artifact and, following successful verification of live human presence through a liveness detection module, produces a hardware-backed signature over the digest using a device-resident secure element. This design cryptographically binds each artifact to a specific instance of verified human presence, producing a persistent, transferable, and publicly verifiable tag while preventing replay across artifacts. Importantly, A-Tag does not perform identification, authorization, or artifact binding to a specific individual; rather, it provides a narrowly scoped but strong guarantee of \emph{human-initiated action with verified liveness}. To the best of our knowledge, A-Tag is the first system to provide verifiable, hardware-backed, and privacy-preserving binding of digital artifacts to live human presence without relying on identity.

To enable independent verification, A-Tag incorporates transparent key registration and logging. Public keys are generated within device hardware and registered through a secure initialization process that ensures integrity and freshness (e.g., via encrypted transport and anti-replay mechanisms), without requiring user identity or prior authentication. Registered keys are periodically committed to an append-only public ledger via Merkle-tree-based anchoring, enabling verifiers to validate key legitimacy without fully trusting the server, through publicly auditable commitments.
We consider a strong adversarial setting encompassing large-scale automation, relay attacks, and physical interaction with devices. Within this model, A-Tag tightly couples valid tag generation with real-time human presence, providing security guarantees against scalable automated or programmatic attacks.

A key component of A-Tag is a device-integrated liveness detection mechanism that verifies real-time human presence during tag generation. While A-Tag is compatible with a range of approaches, in this work we instantiate it using a recent method that leverages on-device sensor signals to provide robust, low-overhead, and non-intrusive liveness guarantees~\cite{alive}. While we defer detailed discussion of this component to Section~\ref{sec:background}, to guide the design and evaluation of A-Tag, we formulate the following research questions:
\\
\textbf{RQ1:} How can \emph{live human presence} be formalized as a security primitive suitable for integration with cryptographic systems?\\
\textbf{RQ2:} How can a verified liveness event be cryptographically bound to a specific digital artifact?\\
\textbf{RQ3:} What guarantees can be achieved against adversaries attempting to generate valid tags through automation, replay of prior artifacts or signals, and physical emulation of human interaction?\\
\textbf{RQ4:} How can such a system remain usable and low-friction while enforcing meaningful security guarantees?\\
\textbf{RQ5:} How can an independently verifiable model be achieved while minimizing reliance on trusted third parties through transparency and public anchoring?\\
These questions guide the design and evaluation of A-Tag. We address them by building a prototype on commodity devices, exploring practical usage scenarios, and evaluating its performance under realistic conditions. Results demonstrate that A-Tag achieves efficient verification and strong robustness in practice. 

The contributions of this paper are six-fold:
$(i)$ We introduce \emph{human-in-the-loop tagging} as a security primitive, establishing verified human presence as a property that can be cryptographically bound to digital artifacts.
$(ii)$ We present the design of A-Tag, a unified system that combines artifact hashing, freshness mechanisms, liveness verification, and hardware-backed signing to enforce human-presence-gated tagging.
$(iii)$ We design a transparency-backed key registration and verification mechanism with periodic public anchoring, enabling trust-minimized verification of device credentials.
$(iv)$ We address key lifecycle management, including revocation through the transparency layer, ensuring that verification accounts for both cryptographic tag validity and current key status.
$(v)$ We formalize the threat model, encompassing large-scale automation, relay attacks, and physical interaction with devices, and analyze A-Tag's security guarantees against liveness-targeted and cryptographic attacks.
$(vi)$ We implement A-Tag on commodity devices and experimentally evaluate its performance, demonstrating deterministic verification, low latency, and practical deployability.

Overall, A-Tag provides a principled, deployable approach to reintroducing verifiable human presence into digital systems, addressing a growing security gap as computation becomes increasingly autonomous.
The remainder of this paper is organized as follows. Section~\ref{sec:background} provides background on liveness detection systems. Section~\ref{sec:system} defines the System Model and Problem. Section~\ref{sec:design} presents the design overview, followed by the A-Tag construction in Section~\ref{sec:proposed}. Section~\ref{sec:analysis} analyzes system security. Section~\ref{sec:implementation} describes the implementation, and Section~\ref{sec:evaluation} presents the experimental evaluation. Section~\ref{sec:related} discusses related work, and Section~\ref{sec:conclusion} concludes.

\section{Background on Liveness Detection}
\label{sec:background}

Liveness detection aims to distinguish live human input from automated or synthetic interactions. This distinction has become increasingly important as AI-generated and automated interactions grow in realism and scale. Existing approaches broadly fall into three categories: challenge-response, appearance-based, and physiological sensing~\cite{pad-survey}. These approaches involve different trade-offs between security, usability, and deployability. Methods that are scalable and low-friction are often more susceptible to adversarial manipulation, whereas those offering stronger guarantees typically require explicit user interaction or specialized hardware.

Within physiological sensing, a promising direction leverages neuromuscular micro-motion signals, which can be captured using commodity inertial sensors~\cite{alive}. Human motor behavior is governed by neural activity and neuromuscular feedback loops, which exhibit inherent variability. Prior work shows that this variability originates in neural control~\cite{yale2015} and propagates to observable motion through neuromuscular coupling~\cite{cortico2015}. Even in near-static conditions, these processes produce small but structured fluctuations, referred to as \emph{micro-motion}, which can be captured and analyzed using inertial measurement units (IMUs) in commodity devices~\cite{livenessPatent}. Such signals exhibit statistical patterns tied to human physiology, making them difficult to reproduce using synthetic or mechanical systems without high-precision actuation.

IMUs, widely available in smartphones and wearables, are sufficiently sensitive to capture these micro-motion signals, making them particularly attractive for practical liveness detection. Unlike approaches requiring explicit challenges or specialized sensors, IMU-based micro-motion enables passive, continuous, and low-friction verification using commodity hardware. Although not impossible to replicate, reproducing these signals in real time and at scale presents significant practical challenges for adversaries. These characteristics make IMU-based micro-motion a suitable choice for A-Tag's liveness mechanism. Building on the feasibility demonstrated in prior work~\cite{alive}, A-Tag adopts this approach not merely as a classification mechanism, but as a \emph{cryptographic gating mechanism} that conditions security-critical signing operations on evidence of human presence.

\section{System Model and Problem Definition}
\label{sec:system}

In this section, we formalize the architecture, entities, protocols, and security assumptions underlying \textit{A-Tag}. The design of A-Tag is guided by RQ1–RQ3, which concern the formalization of human presence and its secure binding to digital artifacts. 
Table~\ref{tab:notation} summarizes the core notation used throughout this section and the rest of the paper.
Given a file $f$, the A-Tag system produces a tag $\tau$ such that the pair $(f, \tau)$ certifies that:
\emph{At tagging time, a cryptographic signing operation over a digest of $f$ was performed on a user-controlled device, and this operation was gated by a successful liveness check indicating the physical presence of a human user.}

Importantly, $\tau$ does not embed raw liveness evidence. Instead, it attests that the signing operation was authorized only following a successful liveness decision, with the signing itself executed within the trusted execution environment. This design minimizes leakage of sensitive sensor data while preserving strong guarantees about the conditions under which the tag was generated.
These guarantees are realized by combining on-device liveness detection, hardware-backed key protection and signing, and a publicly auditable key transparency infrastructure. We next describe the system entities and components, followed by the system workflow, adversary model, and security goals.

\begin{table}[t]
\centering
\caption{Notation used throughout the paper.}
\label{tab:notation}
\begin{tabular}{l l}
\toprule
Symbol & Description \\
\midrule
$f$ & Digital artifact \\
$\tau$ & Tag (attestation token) \\
$H(\cdot)$ & Collision-resistant hash function \\
$h_f$ & Artifact hash, $H(f)$ \\
$\mathsf{cid}$ & Client identifier \\
$\mathsf{n}$ & Client-generated nonce \\
$h_c$ & Client data hash, $H(\mathsf{cid} \parallel \mathsf{n})$ \\
$m_{\text{init/tag}}$ & Message signed during corresponding phase \\
$\mathsf{authData}_{\mathrm{init/tag}}$ & Authenticated metadata for protocol messages\\
$(sk, pk)$ & Private and public key pair \\
$\sigma$ & Digital signature \\
$\mathsf{ctr}$ & Monotonic signing counter \\
$\mathsf{Liveness}()$ & Binary function indicating human presence \\
\bottomrule
\end{tabular}
\end{table}

\subsection{Entities and Components}

\begin{figure}[!t]
    \centering         
    \includegraphics[width=\linewidth]{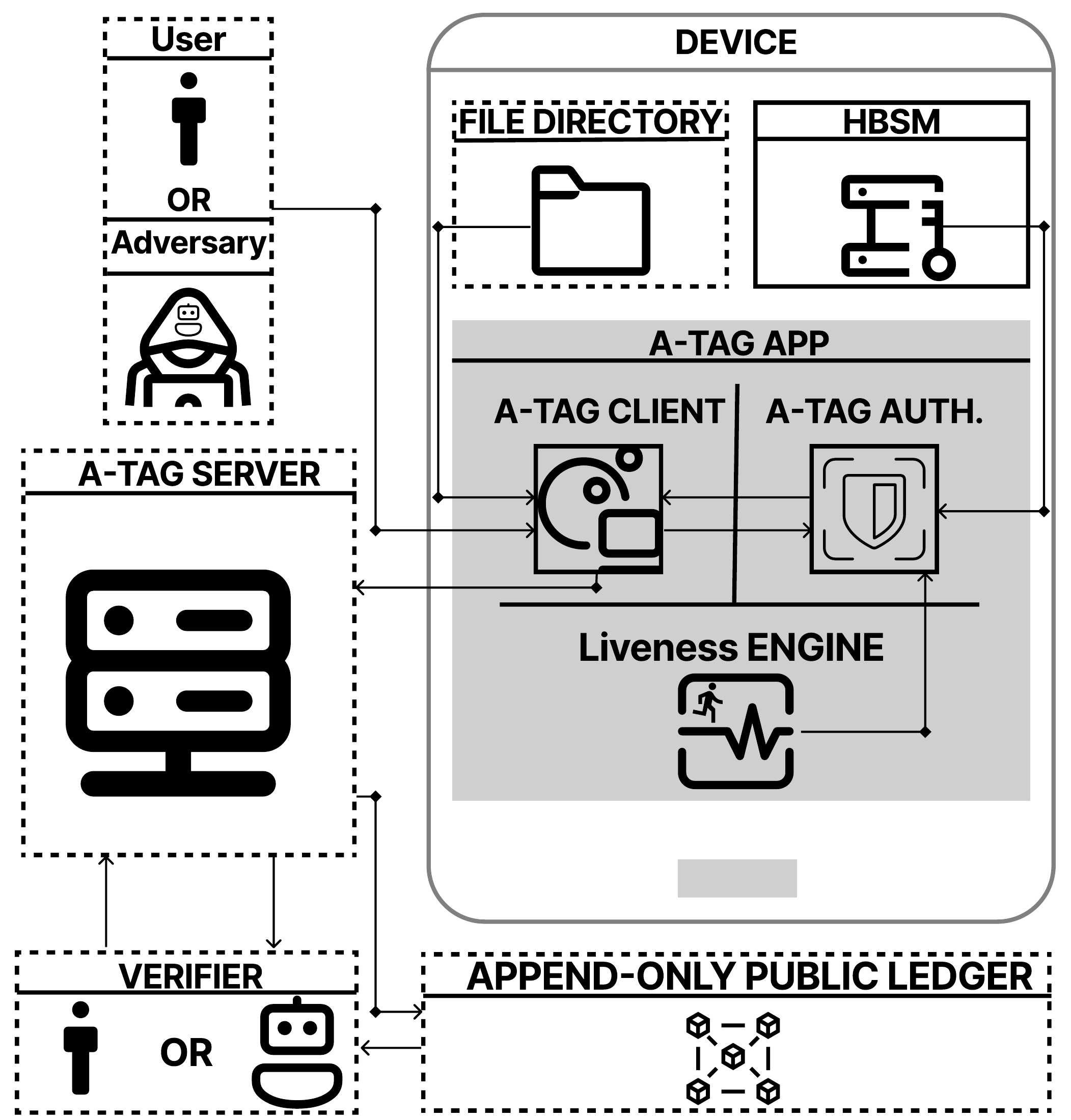}
    \caption{A-Tag system architecture and workflow.}
    \label{fig:architecture}
\end{figure}

Figure~\ref{fig:architecture} illustrates the overall architecture and tagging workflow of A-Tag. The system consists of on-device components responsible for tag generation and external infrastructure that enables public, auditable verification. At a high level, a user or adversary interacts with a client device to request tagging of a file, while trusted on-device components enforce liveness-gated signing. The resulting tag can be verified by any party, either locally or via the transparency service.
The figure also highlights the system’s trust boundaries. In particular, only the hardware-backed security module (HBSM) is assumed to provide a strong root of trust, while other on-device components operate in an untrusted or partially trusted environment. External services, including the A-Tag server and the append-only public ledger, are designed to be publicly auditable. The server is trusted for key registration and revocation, but its actions are transparently anchored in the public ledger, reducing reliance on continuous trust.
We next describe the main entities and components involved in the A-Tag architecture.

\textbf{User.} A human interacting with the system through a client device to generate tags.

\textbf{Adversary.} An entity attempting to generate valid tags without satisfying liveness requirements, potentially controlling software or interacting physically with the device.

\textbf{Client Device.} A user-controlled device (e.g., smartphone or wearable) equipped with motion sensors (e.g., IMU), a secure hardware component (e.g., Trusted Execution Environment or secure element), and the A-Tag application.

\textbf{A-Tag Application.} The A-Tag application is logically composed of three on-device components: the A-Tag client, the A-Tag authenticator, and the Liveness Engine. These components jointly realize the tagging workflow, with each component responsible for a distinct role in the tagging pipeline. We adopt terminology inspired by the FIDO architecture, in particular the Client to Authenticator Protocol (CTAP)~\cite{ctap}, which defines the interaction between a client and an authenticator. In contrast to FIDO, A-Tag does not perform user authentication or identification; instead, it enforces liveness-gated signing. We describe each component below.

\textbf{A-Tag Client.} The user-facing component responsible for initiating tagging requests. It allows a user or adversary to select a target file $f$, computes a digest of $f$, and constructs a request to the A-Tag authenticator.

\textbf{A-Tag Authenticator.} A trusted on-device component that processes tagging requests. Upon receiving a request, it validates the request structure, enforces liveness verification via the Liveness Engine, and, upon success, invokes secure hardware to produce a signature over tag-related data. The A-Tag authenticator mediates all access to the signing key.

\textbf{Liveness Engine.} We model human presence verification as a binary function
$\mathsf{Liveness}() \rightarrow \{0,1\}$, which outputs $1$ if a human user is physically present at the client device at invocation time, and $0$ otherwise.
The selected implementation reports a false acceptance rate (FAR) of 0.00015 across 61,000 attempts on Android and no observed false acceptances across 40,000 attempts on iOS~\cite{alive}.
The Liveness Engine is replaceable, and its underlying detection algorithm is outside the scope of A-Tag.

\textbf{Hardware-Backed Secure Module (HBSM).} A device-integrated, hardware-backed trusted execution environment responsible for secure key management and cryptographic operations. The HBSM generates and stores cryptographic keys in a non-exportable manner and ensures that all signing operations are executed within an isolated environment protected from the host operating system.
A-Tag requires that the signing key be provisioned within an HBSM offering hardware-backed protection (e.g., Android StrongBox or TEE-backed Keystore, or Apple Secure Enclave). The module enforces access control policies such that signing operations are performed only upon authorization from the A-Tag Authenticator, following successful liveness verification.

\textbf{Verifier.} Any party that verifies the validity of a tagged artifact $(f, \tau)$ by recomputing a cryptographic digest of $f$ and validating the tag using the corresponding public key. In A-Tag, public keys are managed externally to the tag and are obtained through a server-assisted lookup, reflecting the requirement that valid keys must be registered and maintained by the A-Tag service, rather than being self-asserted within the tag. 

\textbf{A-Tag Server.} A service that manages public key registration and revocation. The server maintains an append-only log of key states and periodically anchors it to a public ledger to enable external auditability. It also provides proofs required for verification, such as inclusion and revocation proofs.

\textbf{Public Ledger.} A decentralized, publicly verifiable system used as an immutable anchoring layer for commitments generated by the A-Tag Server. By recording periodic commitments to the server’s log, the ledger ensures global consistency and prevents equivocation, allowing any verifier to check that a given view of the key registry is consistent with a publicly committed state.

\subsection{System Workflow}
A-Tag operates through three logically distinct procedures: initialization, tag generation, and verification. Initialization provisions a device-bound key under verified human presence. Tag generation uses this key to produce tags on demand. Verification is decoupled from these operations and can be performed by any party.

\textbf{Initialization and Key Registration.}
During initialization, the client device generates a key pair $(sk, pk)$ within the HBSM in the presence of a live human user. The private key $sk$ remains non-exportable, while the public key $pk$ is registered with the A-Tag server. This process may be repeated for key rotation or multiple devices. The server maintains an append-only log of registered public keys and their status (e.g., active or revoked), periodically anchoring its state to a public ledger and providing proofs of inclusion and revocation. This enables external verification of key validity while reducing reliance on trust in the server.

\textbf{Tag Generation.}
To tag a file $f$, the A-Tag client computes its digest $h_f = H(f)$ and forwards a request to the A-Tag Authenticator, which validates its structure and invokes the Liveness Engine. A failed liveness check aborts the operation. Upon successful verification, the authenticator requests the HBSM to sign data derived from $h_f$ using the device-bound private key $sk$. The resulting signature $\sigma$ is combined with the signed data to form the tag $\tau$, producing the tagged artifact $(f, \tau)$. The tag attests to the use of a registered device-bound key under liveness-gated execution without embedding raw liveness evidence, auxiliary sensor data, or the corresponding public key.

\textbf{Tag Verification.}
Given $(f, \tau)$, a verifier recomputes the file digest, checks consistency with the signed content, and verifies the signature using the corresponding public key. The verifier also obtains server transparency data and checks it against commitments anchored in the public ledger to validate the key's registration and revocation status, minimizing server trust. A tag is valid if (i) the file digest matches the signed content, (ii) the public key is registered and unrevoked, and (iii) the signature is correct.

\subsection{Adversary Model and Assumptions}
\label{sec:assumption}

A-Tag operates under a bounded adversarial model with explicitly defined trust assumptions. Its security relies on the correct execution of the on-device A-Tag Authenticator and Liveness Engine, together with hardware-backed protection of cryptographic keys.

\textbf{Adversary Model.}
We consider a probabilistic polynomial-time (PPT) adversary $\mathcal{A}$ who can submit arbitrary tagging requests, replay observed protocol messages, and physically interact with the client device to influence sensor inputs. The adversary may attempt to trigger signing operations without human presence, manipulate physical sensor inputs, or reuse previously observed protocol responses. However, we assume that the adversary cannot compromise or modify the trusted A-Tag Authenticator or Liveness Engine, or bypass their execution during the signing process. The adversary also cannot extract private keys from the HBSM or compromise its hardware-enforced key isolation and access-control mechanisms.

\textbf{Trusted Components and Assumptions.}
A-Tag relies on the following trusted components and assumptions:\\
\textbf{- A-Tag Authenticator and Liveness Engine.} The A-Tag Authenticator and Liveness Engine are trusted on-device components responsible for enforcing the liveness-gated signing policy. Their code and execution are assumed to remain protected from adversarial modification or bypass. The Authenticator invokes the Liveness Engine for every signing request and proceeds with signing only after successful verification. Each liveness check is fresh and bound to the corresponding signing request; a previous successful result cannot be reused to authorize another operation.\\
\textbf{- Hardware-Backed Secure Module.} The HBSM is trusted to enforce key isolation, non-exportability, and hardware-backed cryptographic execution. All signing keys are generated within and remain bound to the HBSM. It protects the private keys and enforces its configured platform-level access-control policies. The A-Tag Authenticator, rather than the HBSM, enforces the application-level liveness requirement before invoking the signing operation.\\
\textbf{- Sensor Input Integrity.} The sensor path used by the Liveness Engine is trusted against software-level injection of arbitrary digital data. Adversaries may influence physical sensor inputs, but cannot directly replace sensor readings with arbitrary digitally injected signals. The effectiveness of liveness detection against physical manipulation and other sensor-level attacks is subject to the capabilities and limitations of the implemented detection mechanism.\\
\textbf{- Transparency Log.} The transparency log is assumed to be append-only and publicly auditable, with equivocation detectable by independent observers.\\
\textbf{- Cryptographic Primitives.} We assume collision-resistant hashing and digital signatures that are existentially unforgeable under chosen-message attacks (EUF-CMA).

\textbf{Formal Security Definition.}
We formalize the core guarantee of A-Tag as \emph{liveness-gated unforgeability}. Let $\Pi$ denote the A-Tag scheme with algorithms $(\mathsf{KeyGen}, \mathsf{Sign}, \mathsf{Verify})$. The security experiment proceeds as follows. The challenger $\mathcal{C}$ generates a key pair $(sk, pk) \leftarrow \mathsf{KeyGen}$ and provides the adversary $\mathcal{A}$ with access to an oracle $\mathcal{O}_{\text{tag}}(\cdot)$ modeling the trusted A-Tag signing interface. On input a message $x$, the oracle performs a fresh liveness evaluation $b \leftarrow \mathsf{Liveness}()$ and returns $\sigma \leftarrow \mathsf{Sign}(sk,x)$ if $b=1$, and $\bot$ otherwise. The oracle models the enforced signing path: the adversary cannot access the underlying signing operation independently of this interface.
Eventually, $\mathcal{A}$ outputs a candidate forgery $(x^*,\sigma^*)$. The adversary wins if
$
\mathsf{Verify}(pk,x^*,\sigma^*)=1
$
and $(x^*,\sigma^*)$ was not returned by a successful oracle query. The advantage of $\mathcal{A}$ is defined as
$
\mathsf{Adv}_{\Pi}(\mathcal{A})
=
\Pr[\mathcal{A}\text{ wins}].
$
We say that A-Tag satisfies liveness-gated unforgeability if for all PPT adversaries $\mathcal{A}$,
$
\mathsf{Adv}_{\Pi}(\mathcal{A})\leq \epsilon(\lambda),
$
where $\epsilon(\lambda)$ is negligible in the security parameter $\lambda$.

This definition captures the unforgeability of signatures produced outside the authorized signing interface, under the assumption that the trusted Authenticator cannot be bypassed and that every invocation of the interface requires a fresh liveness verification. It does not establish resistance to compromise of the trusted Authenticator, Liveness Engine, or their execution environment.

\subsection{Security Goals and Non-Goals}
\label{sec:securityGoal}

A-Tag cryptographically binds digital artifacts to signing events authorized through fresh human-presence verification. It combines hardware-backed key protection with transparency-backed registration and revocation to enable public verifiability while minimizing trust in the A-Tag server. Liveness serves solely as a gating condition enforced by the trusted A-Tag Authenticator before HBSM signing and is not embedded in the resulting tag.

\textbf{Security Goals.}
A-Tag aims to provide the following:\\
\textbf{- Unforgeability.}
It is computationally infeasible to generate a valid tag $\tau$ without access to the device-bound private key through the authorized A-Tag signing interface, which is mediated by the trusted A-Tag Authenticator, after successful liveness verification.\\
\textbf{Artifact Binding.}
Each tag is cryptographically bound to a specific artifact digest $h_f$. Any modification to the artifact results in verification failure.\\
\textbf{Human Presence Binding.}
Tag generation is gated by a fresh liveness check performed by the Liveness Engine and enforced by the A-Tag Authenticator before each signing operation. This ensures that, under the trusted-component assumptions, each newly generated valid tag corresponds to a human-presence verification.\\
\textbf{Replay Resistance.}
The system prevents reuse of previously obtained signing responses as authorization for new tagging operations. Each signing request requires a fresh liveness verification, and the signing process incorporates a per-key monotonic counter maintained within the HBSM to distinguish signing operations.\\
\textbf{Public Verifiability.}
Any party can verify the validity of a tag using publicly available information, with mechanisms to validate key registration and revocation against transparency commitments anchored in a public ledger.\\
\textbf{Transparency and Accountability.}
All credential lifecycle events are recorded in an append-only transparency log, enabling independent auditing of key registration and revocation.\\
\textbf{Revocation Security.}
Compromised credentials can be revoked and the revoked credentials are rejected during tag verification.

\textbf{Non-Goals.}
A-Tag does not aim to provide the following:\\
\textbf{User Identity.}
Authentication or identification of the individual operating the device.\\
\textbf{Content Awareness.}
Evidence that the user reviewed or understood the tagged artifact.\\
\textbf{Human Content Creation.}
A guarantee that the artifact itself was generated by a human.\\
\textbf{Temporal Attestation.}
A-Tag does not provide trusted wall-clock timestamps or independently attest to the actual time of a tagging operation. Although the signing process maintains a per-key monotonic counter, this counter does not establish an externally verifiable timestamp or absolute temporal ordering across different devices. Tags provide content-integrity verification without establishing when the signing event occurred.

\section{Design Overview}
\label{sec:design}
While Section~\ref{sec:system} defines the system model, it does not capture A-Tag’s key design principle. This section distills that principle and shows how A-Tag realizes artifact-bound human presence through liveness verification, hardware-backed signing, and transparency-backed key management.

\textbf{Design Principles.}
A-Tag addresses a missing security property, \emph{artifact-bound human presence}, not provided by existing authentication or signing systems. Informally, this property requires that every valid tag corresponds to an execution event in which a live human authorizes signing of a specific artifact.
A-Tag realizes this property through three constraints. First, \emph{liveness-gated signing} conditions access to the signing key on real-time human presence. Second, \emph{cryptographic artifact binding} requires each signature to be computed over an artifact digest, preventing reuse across content. Third, \emph{hardware-enforced key protection} confines private keys and signing policies within an HBSM, ensuring non-exportability and controlled execution. These constraints are supported by a transparency-backed key management infrastructure, where registration and revocation events are recorded in an append-only log, enabling public verifiability.

At a high level, A-Tag couples three components typically deployed independently: liveness detection, hardware-backed signing, and publicly auditable key management. Given a tagging request for file $f$, the system binds subsequent operations to the artifact digest $h_f$. Any modification to $f$ changes the digest and invalidates the tag. Before signing, a liveness check gates access to the signing key within the HBSM. Upon successful verification, the HBSM signs a message containing $h_f$. Since the key is non-exportable and signing is mediated by the trusted execution environment, valid signatures can only be produced through liveness-gated execution.
A-Tag avoids embedding explicit timing or freshness metadata in tags. Instead, validity relies on artifact integrity and the current status of the signing key, with revocation enforced by the A-Tag server and publicly auditable via the transparency-backed key management service. The resulting tag $\tau$ encapsulates the signed digest and serves as a portable attestation that the artifact was processed under these conditions. Verification requires only the artifact and the corresponding public key from the transparency-backed registry.

\textbf{Security Intuition.}
The security of A-Tag arises not from its individual components, but from their constrained composition. In particular, A-Tag eliminates the gap between human verification and cryptographic action present in prior approaches.
A naive combination of liveness-enabled authentication and cryptographic signing does not provide artifact-bound human presence. For example, a user may authenticate using a biometric mechanism with liveness detection and subsequently sign arbitrary content using the unlocked device. Although liveness was verified during authentication, the signing operation itself is not guaranteed to occur under live human interaction. Recent work~\cite{sp23} further demonstrates that widely deployed face recognition systems remain vulnerable to low-cost physical spoofing attacks, highlighting the risk of decoupling liveness verification from the protected cryptographic operation.

A-Tag prevents this separation by enforcing a non-transferable coupling among: (i) the artifact being processed, (ii) the execution context performing the signing, and (iii) the liveness-gated release of signing capability. Artifact binding ensures that signatures are specific to a given file. Liveness gating within the trusted execution environment ensures that signing requires a fresh human presence check and cannot be replayed or invoked out of context. Finally, non-exportable keys and hardware-mediated signing prevent external access or reuse of signing capability.

A-Tag achieves these guarantees without embedding explicit freshness or timing information into tags. While freshness is enforced during initialization, tag generation relies on the enforced coupling of artifact binding and liveness-gated execution rather than per-tag timestamps or nonces. Consequently, previous signatures cannot be transferred to other artifacts, and producing a new valid signature requires a new liveness-verified interaction.
Thus, a valid tag provides evidence of a concrete execution event in which a live human authorized signing of a specific artifact under hardware-enforced constraints. This property emerges from the composition of human presence verification, artifact binding, hardware protection, and public verifiability rather than from any individual mechanism alone. We formalize tag construction and the associated cryptographic protocols in Section~\ref{sec:proposed}.

\section{A-Tag Construction}
\label{sec:proposed}

In this section, we formalize A-Tag as a set of protocols realizing human-bound, artifact-specific attestation. The system operates in three phases: (i) initialization, where cryptographic keys are provisioned and registered; (ii) tag generation, where a live human authorizes signing of a specific artifact; and (iii) verification, where any party validates a tag using publicly auditable information without fully trusted intermediaries.
We model A-Tag using standard cryptographic primitives. Let $H(\cdot)$ denote a collision-resistant hash function and $(\mathsf{KeyGen}, \mathsf{Sign}, \mathsf{Verify})$ a digital signature scheme, where $\mathsf{KeyGen}$ outputs $(sk,pk)$, $\mathsf{Sign}(sk,m)$ outputs signature $\sigma$, and $\mathsf{Verify}(pk,m,\sigma)$ outputs $1$ for valid signatures and $0$ otherwise. Private key operations execute within a secure module, and signing access is mediated by the A-Tag authenticator under liveness-gated control. We model $\mathsf{Liveness}()$ as an oracle that cannot be satisfied by purely automated processes and resists replay or precomputation.

\textbf{Initialization and Key Registration:}
A-Tag requires one-time initialization per application instance to provision a cryptographic key bound to a human-controlled device. The same procedure applies to key rotation and device migration. The A-Tag client orchestrates the protocol by obtaining an attested public key from the authenticator and registering it with the server. We describe this protocol from the client, authenticator, and server perspectives.

\textit{Client Initialization.}
The client generates a fresh identifier $\mathsf{cid} \leftarrow \{0,1\}^{256}$ and a nonce $\mathsf{n} \leftarrow \{0,1\}^{128}$, and computes $h_c = H(\mathsf{cid} \parallel \mathsf{n})$ and $h_{\mathsf{cid}} = H(\mathsf{cid})$. The client then constructs a credential creation request containing $h_c$ and $h_{\mathsf{cid}}$, along with the desired cryptographic parameters, and sends it to the A-Tag authenticator.

\textit{Authenticator Processing and Key Generation.}
Upon receiving the request, the A-Tag authenticator evaluates $\mathsf{Liveness}()$. If $\mathsf{Liveness}() = 1$, it invokes the HBSM to generate a key pair $(sk, pk) \leftarrow \mathsf{KeyGen}()$, where $sk$ remains non-exportable. The authenticator then constructs
$\mathsf{authData}_{\text{init}} = (h_{\mathsf{cid}}, \mathsf{flag}, \mathsf{ctr}, pk)$,
where $\mathsf{ctr}$ is a per-key monotonic counter maintained within the HBSM and incremented only on each successful signing operation (with $\mathsf{ctr} = 0$ at initialization). It then forms $m_{\text{init}} = (h_c \parallel \mathsf{authData}_{\text{init}})$ and computes $\sigma \leftarrow \mathsf{Sign}(sk, m_{\text{init}})$ within the HBSM, returning the result to the client. 
The A-Tag application verifies that the generated key is hardware-backed and aborts registration if this requirement is not satisfied. The authenticator also obtains hardware attestation evidence $\mathsf{att}$ associated with $pk$, which is returned to the client along with the signed initialization data.

\textit{Client Verification and Registration.}
The authenticator returns to the client the tuple $(\sigma, \mathsf{authData}_{\text{init}})$. 
Upon receiving the response, the client parses $\mathsf{authData}_{\text{init}}$ and verifies correctness by checking:
(i) consistency of $h_c$ with the original request,
(ii) consistency of $h_{\mathsf{cid}}$ with the locally computed value,
(iii) validity of the user presence flag,
(iv) correct initialization of the counter $\mathsf{ctr}$, and
(v) validity of the signature $\sigma$ under the public key $pk$ over the reconstructed message $(h_c \parallel \mathsf{authData}_{\text{init}})$,
$\mathsf{Verify}(pk, h_c \parallel \mathsf{authData}_{\text{init}}, \sigma) \stackrel{?}{=} 1$.
Note that successful signature verification implicitly authenticates the integrity and origin of all fields within $\mathsf{authData}_{\text{init}}$, including the public key $pk$. Hardware attestation evidence is separately used to establish the hardware origin of $pk$.

\begin{algorithm}
\caption{A-Tag Initialization and Key Registration}
\label{alg:atag_init}
\begin{algorithmic}

\STATE{\textbf{Client:}}
\STATE $\mathsf{cid} \leftarrow \{0,1\}^{256}$,\quad $\mathsf{n} \leftarrow \{0,1\}^{128}$
\STATE $h_c \leftarrow H(\mathsf{cid} \parallel \mathsf{n})$,\quad $h_{\mathsf{cid}} \leftarrow H(\mathsf{cid})$
\STATE send $(h_c, h_{\mathsf{cid}})$ to authenticator

\vspace{0.5\baselineskip}

\STATE{\textbf{Authenticator / HBSM:}}
\IF{$\mathsf{Liveness}() = 0$} \STATE reject \ENDIF
\STATE $(sk, pk) \leftarrow \mathsf{KeyGen}()$
\STATE initialize per-key counter $\mathsf{ctr} \leftarrow 0$
\STATE $\mathsf{authData}_{\text{init}} \leftarrow (h_{\mathsf{cid}}, 1, \mathsf{ctr}, pk)$
\STATE $\sigma \leftarrow \mathsf{Sign}(sk,\, h_c \parallel \mathsf{authData}_{\text{init}})$
\STATE return $(\sigma, \mathsf{authData}_{\text{init}})$

\vspace{0.5\baselineskip}

\STATE{\textbf{Client:}}
\IF{$\mathsf{Verify}(pk,\, h_c \parallel \mathsf{authData}_{\text{init}}, \sigma) \neq 1$}
    \STATE reject
\ENDIF
\STATE send $(h_{\mathsf{cid}}, pk, \mathsf{att})$ to server

\vspace{0.5\baselineskip}

\STATE{\textbf{Server:}}
\STATE append $(h_{\mathsf{cid}}, pk)$ to log
\STATE update and commit accumulator

\end{algorithmic}
\end{algorithm}

\begin{algorithm}
\caption{A-Tag Tag Generation}
\label{alg:atag_tagging}
\begin{algorithmic}

\STATE{\textbf{Client:}}
\STATE $h_f \leftarrow H(f)$,\quad $h_{\mathsf{cid}} \leftarrow H(\mathsf{cid})$
\STATE send $(h_f, h_{\mathsf{cid}}, 1)$ to authenticator

\vspace{0.5\baselineskip}

\STATE{\textbf{Authenticator / HBSM:}}
\IF{$\mathsf{Liveness}() = 0$} \STATE reject \ENDIF
\STATE retrieve $sk$ using $h_{\mathsf{cid}}$
\STATE increment per-key counter $\mathsf{ctr}$
\STATE $\mathsf{authData}_{\text{tag}} \leftarrow (h_{\mathsf{cid}}, 1, \mathsf{ctr})$
\STATE $\sigma \leftarrow \mathsf{Sign}(sk,\, h_f \parallel \mathsf{authData}_{\text{tag}})$
\STATE return $(\sigma, \mathsf{authData}_{\text{tag}})$

\vspace{0.5\baselineskip}

\STATE{\textbf{Client:}}
\IF{$\mathsf{Verify}(pk,\, h_f \parallel \mathsf{authData}_{\text{tag}}, \sigma) \neq 1$}
    \STATE reject
\ENDIF
\STATE $\tau \leftarrow (h_f, \mathsf{authData}_{\text{tag}}, \sigma)$

\end{algorithmic}
\end{algorithm}
If all checks succeed, the client submits $(h_{\mathsf{cid}}, pk, \mathsf{att})$ to the A-Tag server over an authenticated, integrity-protected channel. The server verifies the application's integrity and validates $\mathsf{att}$ to confirm that $pk$ is hardware-backed. Registration is rejected if either check fails; otherwise, $pk$ is added to an append-only log indexed by $h_{\mathsf{cid}}$, with its validity status maintained as follows.
We model the server state using two sets: $\mathcal{K}_{\mathrm{reg}}$ for registered keys and $\mathcal{K}_{\mathrm{rev}}$ for revoked keys. A key is valid if registered and not revoked:
$\mathsf{ValidKey}(pk) := (pk \in \mathcal{K}_{\mathrm{reg}}) \land (pk \notin \mathcal{K}_{\mathrm{rev}})$.
For public auditability, the server implements these sets using append-only Merkle trees and periodically anchors their state to a public ledger, enabling external parties to verify the consistency.
Successful registration completes initialization, establishing a hardware-protected signing key bound to an application instance. Algorithm~\ref{alg:atag_init} summarizes the initialization and key registration workflow.

\textbf{Tag Generation:}
The tag generation protocol binds a digital artifact to a liveness-gated signing event enforced by the A-Tag authenticator and executed within the HBSM. It proceeds in three stages: client request, liveness-gated signing, and client-side verification and tag formation, each described below.

\textit{Client Request.}
Given an artifact $f$, the client computes $h_f = H(f)$ and $h_{\mathsf{cid}} = H(\mathsf{cid})$, then sends $(h_f, h_{\mathsf{cid}}, \mathsf{flag})$ to the authenticator, where $\mathsf{flag}$ indicates user presence intent. 

\textit{Liveness-Gated Signing.}
Upon receiving the request, the authenticator first enforces a configurable rate limit on tagging operations, set to 10 requests per minute in our implementation. If the request is within the allowed limit, the authenticator evaluates $\mathsf{Liveness}()$. If successful, it retrieves the corresponding private key via $h_{\mathsf{cid}}$ and invokes the HBSM to sign. The authenticator constructs
$\mathsf{authData}_{\text{tag}} = (h_{\mathsf{cid}}, \mathsf{flag}, \mathsf{ctr})$. The signed message is
$m_{\text{tag}} = (h_f \parallel \mathsf{authData}_{\text{tag}})$,
and the HBSM outputs
$\sigma \leftarrow \mathsf{Sign}(sk, m_{\text{tag}})$.

\textit{Client Verification and Tag Formation.}
The authenticator returns $(\sigma, \mathsf{authData}_{\text{tag}})$. The client verifies consistency of $h_{\mathsf{cid}}$, validity of the user presence flag, monotonicity of $\mathsf{ctr}$, and correctness of the signature
$\mathsf{Verify}(pk, h_f \parallel \mathsf{authData}_{\text{tag}}, \sigma) \stackrel{?}{=} 1$.
If valid, the client outputs the tag
$\tau = (h_f, \mathsf{authData}_{\text{tag}}, \sigma).$
Algorithm~\ref{alg:atag_tagging} summarizes the tag generation workflow, including artifact hashing, liveness-gated signing within the secure module, and client-side verification prior to tag construction.

\textbf{Verification:}
The verification procedure enables any party to validate that an artifact is associated with a valid tag, without requiring access to trusted intermediaries beyond publicly verifiable key material. Verification ensures that the artifact matches the tag and that the signing key is legitimate under the transparency-backed registration system.
This subsection reviews the phases of tag verification—\textit{tag extraction and artifact processing}, \textit{public key resolution and transparency}, and \textit{signature verification}—and presents the corresponding pseudocode.

\textit{Tag Extraction and Artifact Processing.}
Given an artifact $f$ and tag 
$\tau = (h_f, \mathsf{authData}_{\text{tag}}, \sigma)$, 
the verifier recomputes
$h_f' = H(f)$
and verifies that
$h_f' \stackrel{?}{=} h_f$.
If the equality fails, the verifier rejects.
Next, the verifier parses $\mathsf{authData}_{\text{tag}}$ as 
$(h_{\mathsf{cid}}, \mathsf{flag}, \mathsf{ctr})$.
The verifier subsequently reconstructs the signed message as
$m_{\text{tag}} = (h_f \parallel \mathsf{authData}_{\text{tag}})$,
which is used for signature verification.

\textit{Public Key Resolution and Transparency.}
To verify the signature, the verifier first obtains the public key $pk$ associated with $h_{\mathsf{cid}}$. If the verifier corresponds to the originating client, $pk$ is retrieved locally; otherwise, the verifier queries the A-Tag server. If no record is found, the verifier rejects. Regardless of how $pk$ is obtained, the verifier must query the A-Tag server to establish that the key remains registered and has not been revoked.
Optionally, the verifier may independently audit the server's response using the blockchain-based transparency mechanism. Specifically, the verifier checks whether the server-reported key status is consistent with the latest publicly committed log state, including whether $pk$ is registered and has not been revoked. Failure of this audit results in rejection.
This separation ensures that signature verification depends only on successful key resolution, while transparency provides an additional, publicly verifiable consistency check on the server-maintained state.

\textit{Signature Verification.}
The verifier accepts $\tau$ iff (C1) $h_f' = h_f$, (C2) $pk$ is valid and non-revoked, and (C3) $\mathsf{Verify}(pk, m_{\text{tag}}, \sigma) = 1$. A valid tag thus provides verifiable evidence—under the stated assumptions—that the artifact was cryptographically bound to a liveness-gated signing event using a registered, hardware-protected key.
Condition (C2) relies on the mandatory server-side key-status check, while optional blockchain auditing enables independent verification against publicly committed log states. Accordingly, verification relies on the A-Tag server for current key-status information, assuming an honest-but-auditable server and a trusted secure module.
The monotonic counter is not evaluated during verification because A-Tag does not maintain or expose tag state externally. Instead, it enforces ordering and detects anomalies within the trusted client–authenticator interaction without introducing linkability or server-side tag tracking.
Algorithm~\ref{alg:atag_verification} summarizes the verification procedure, including artifact validation, public key resolution, transparency checks, and signature verification.

\textbf{End-to-End Security:}
A-Tag binds a human, a device-protected key, and the artifact under a publicly auditable model. During initialization, a non-exportable key is generated and registered in a transparency-backed log. During tag generation, signing is gated by liveness, ensuring that each signature reflects a fresh human-authorized event over a specific artifact digest. Verification ensures that the artifact matches the signed content and that the key is valid under the transparency system. 
Under a trusted secure module and an honest-but-auditable log, A-Tag provides strong guarantees of artifact integrity, key-bound provenance, and accountability, while reducing reliance on intermediaries to verifiable consistency checks.

\begin{algorithm}
\caption{A-Tag Verification}
\label{alg:atag_verification}
\begin{algorithmic}

\STATE{parse $\tau$; compute $h_f' \leftarrow H(f)$; reject if $h_f' \neq h_f$}

\STATE{parse $\mathsf{authData}_{\text{tag}}$; reject if $\mathsf{flag} \neq 1$}

\STATE{$m_{\text{tag}} \leftarrow (h_f \parallel \mathsf{authData}_{\text{tag}})$}

\medskip
\IF{local credential}
    \STATE{retrieve $pk$}
\ELSE
    \STATE{query server with $h_{\mathsf{cid}}$; receive $pk$}
    \IF{no record}
        \STATE{reject}
    \ENDIF
\ENDIF

\medskip
\STATE{query server for current registration and revocation status of $pk$}
\IF{key is not registered or is revoked or its status cannot be established}
    \STATE{reject}
\ENDIF

\medskip
\STATE{optionally audit key status against public log state}
\IF{audit fails}
    \STATE{reject}
\ENDIF

\medskip
\IF{$\mathsf{Verify}(pk, m_{\text{tag}}, \sigma) \neq 1$}
    \STATE{reject}
\ENDIF

\STATE{accept}

\end{algorithmic}
\end{algorithm}

\section{Security Analysis}
\label{sec:analysis}

We analyze A-Tag's security regarding RQ3 under the adversarial model and assumptions defined in Section~\ref{sec:assumption}. We show that A-Tag achieves unforgeability, human presence binding, replay resistance, and strong artifact binding and integrity. These guarantees are verifiable using auditable information without requiring trust in the A-Tag server beyond correct log operation, while maintaining accountability through transparency and revocation mechanisms.

\textbf{Unforgeability:}
A-Tag ensures valid tags are produced only via the HBSM-protected signing interface. Given signing key non-exportability and enforced access control, any successful forgery reduces to breaking the underlying signature scheme or violating HBSM enforcement. The liveness mechanism constitutes an orthogonal access-control assumption governing authorized interface use and is not part of the cryptographic unforgeability guarantee.

\begin{theorem}
\label{thrm:Unforgeability}
Assume (i) $(\mathsf{KeyGen}, \mathsf{Sign}, \mathsf{Verify})$ is EUF-CMA secure, and (ii) the HBSM enforces $\mathsf{sk}$ non-exportability and correct execution of the signing interface. Then A-Tag is unforgeable: no PPT adversary $\mathcal{A}$ can produce a valid tag $\tau = (h_f, \mathsf{authData}_{\text{tag}}, \sigma)$ for a message $m_{\text{tag}} = (h_f \parallel \mathsf{authData}_{\text{tag}})$ not previously signed by the HBSM, except with negligible probability.
\end{theorem}

\begin{proof}
We reduce to the EUF-CMA security of the signature scheme. Let $\mathcal{A}$ be an adversary that outputs a valid forgery $(m_{\text{tag}}^*, \sigma^*)$ such that $m_{\text{tag}}^*$ was not previously queried to the signing interface.
We construct a simulator $\mathcal{B}$ that is given $\mathsf{pk}$ and oracle access to $\mathcal{O}_{\mathsf{Sign}}(\cdot)$. $\mathcal{B}$ simulates the A-Tag signing interface by forwarding each query $m_{\text{tag}}$ from $\mathcal{A}$ to $\mathcal{O}_{\mathsf{Sign}}$ and returning the resulting signature.
When $\mathcal{A}$ outputs $(m_{\text{tag}}^*, \sigma^*)$ such that $\mathsf{Verify}(\mathsf{pk}, m_{\text{tag}}^*, \sigma^*) = 1$, and $m_{\text{tag}}^*$ was not queried to the oracle, $\mathcal{B}$ outputs this pair as a valid EUF-CMA forgery.
Thus, $\mathcal{B}$ succeeds whenever $\mathcal{A}$ succeeds. By EUF-CMA security, this occurs only with negligible probability. Hence, A-Tag is unforgeable. Therefore, A-Tag is unforgeable.
\end{proof}

\textbf{Human Presence Binding:}
A-Tag ensures that each valid tag is produced via a signing interface gated by a fresh liveness-verified interaction. This interface can be modeled as a restricted signing oracle conditioned on successful liveness verification.
\begin{lemma}
\label{lemma:binding}
Assuming the liveness module correctly distinguishes live human interaction from non-human or replayed inputs within its operational limits, any valid tag $\tau = (h_f, \mathsf{authData}_{\text{tag}}, \sigma)$ 
implies that the corresponding signing operation was preceded by a successful and fresh liveness verification classified as human, except with the module's error probability.
\end{lemma}

\begin{proof}
We model the HBSM interface as a signing oracle accepting queries only when accompanied by a fresh liveness verification. By construction, this condition is enforced by the HBSM and cannot be bypassed under the trust assumptions, and $\mathsf{sk}$ is non-exportable.
Therefore, any valid signature must originate from an oracle invocation satisfying the liveness condition. The claim follows from the correctness of the liveness module, up to its error probability.
\end{proof}

\textbf{Replay Resistance:}
A-Tag prevents replays from being interpreted as fresh authorization events by enforcing freshness at two levels: initialization requests bind to client nonces, and tagging operations bind to a hardware-enforced monotonic counter. We model signing as a restricted oracle enforcing these freshness constraints.

\begin{lemma}
\label{lemma:replay}
Let $\mathcal{O}_{\mathsf{Sign}}(\cdot)$ be a signing oracle enforcing (i) nonce freshness during initialization, (ii) strictly monotonic counter progression during tagging. Assuming EUF-CMA security, no PPT adversary can cause a previously generated transcript or tag to be accepted as a fresh authorization event, except with negligible probability.
\end{lemma}

\begin{proof}
We model the adversary $\mathcal{A}$ with access to $\mathcal{O}_{\mathsf{Sign}}(\cdot)$.
\textbf{Initialization.}
Each initialization message includes a client nonce that is incorporated into the signed message. By EUF-CMA security, the adversary cannot modify any field without invalidating the signature, and thus can only replay an identical transcript. However, reuse of a nonce or counter value is detectable and rejected by the server. Therefore, replayed initialization transcripts cannot be accepted as fresh events.\\
\textbf{Tagging.}
Each tag includes a counter value $\mathsf{ctr}$ that is initialized and strictly incremented by the HBSM. Replaying a tag reproduces the same $(h_f, \mathsf{authData}_{\text{tag}})$ and does not yield a fresh authorization. Producing a tag with a new counter or for a different artifact requires a new oracle invocation. By the oracle constraints, this corresponds to a fresh, liveness-gated signing event.
Thus, replay does not enable the adversary to create valid authorization events, completing the proof.
\end{proof}

\textbf{Artifact Binding and Integrity:}
A-Tag binds each tag to a specific artifact by including its digest in the signed message. Any deviation from the original artifact results in verification failure.

\begin{lemma}
\label{lemma:integrity}
Assume (i) the hash function to compute $h_f$ is second-preimage resistant, and (ii) the signature scheme is EUF-CMA secure. Then no PPT adversary can produce a valid tag $\tau = (h_f, \mathsf{authData}_{\text{tag}}, \sigma)$ that verifies for an artifact different from the one originally authorized, except with negligible probability.
\end{lemma}

\begin{proof}
We model the adversary $\mathcal{A}$ with access to the signing interface. Let $\tau$ be a valid tag for artifact $f$. For $\tau$ to verify for a different artifact $f' \neq f$, one of the following must hold:
(i) $h_{f'} = h_f$, violating second-preimage resistance of the hash function, or  
(ii) $\mathcal{A}$ produces a valid signature on a message containing $h_{f'}$, constituting a forgery under EUF-CMA security.
Both cases occur with negligible probability. Therefore, tags are bound to the original artifacts.
\end{proof}

\textbf{Public Verifiability, Transparency, and Revocation:}
A-Tag enables public verification via a transparency log offering verifiable key resolution and recording registration and revocation events. 

\begin{lemma}
\label{lemma:accountability}
Let \(\mathcal{O}_{\mathsf{Log}}(\cdot)\) return sound inclusion and consistency proofs for an append-only log. Assuming EUF-CMA security, any valid tag is verifiable using only public information and \(\mathcal{O}_{\mathsf{Log}}(\cdot)\); no adversary can cause acceptance under an incorrect key except with negligible probability; and any revoked key use is publicly detectable.
\end{lemma}

\begin{proof}
we model \(\mathcal{A}\) with access to \(\mathcal{O}_{\mathsf{Log}}(\cdot)\). Verification uses \(\mathsf{pk}\) and an inclusion proof from the log, which is publicly verifiable, so verification correctness is independent of server trust and depends on the soundness of the transparency log. Acceptance under an incorrect key requires either forging an inclusion proof or a signature, violating log soundness or EUF-CMA security, respectively. Finally, since revocation entries are append-only and globally consistent, hiding a revocation is detectable via inconsistent log views.
\end{proof}

\textbf{Attack Surface and Limitations:}
Table~\ref{tab:attack-analysis} summarizes representative attacks against A-Tag, each of which reduces to violating an underlying assumption or a formally established property (Theorem~\ref{thrm:Unforgeability} and Lemmas~\ref{lemma:binding}, \ref{lemma:replay}, \ref{lemma:integrity}, and \ref{lemma:accountability}), showing that the system introduces no additional attack vectors beyond its trust assumptions. A valid tag attests only that a hardware-protected key was used following a liveness signal classified as human at a specific time, and does not capture user identity, intent, or content correctness; such semantics must be enforced at the application layer.
\begin{table}[t]
\begin{center}
\caption{Attack surface analysis of A-Tag. }
\label{tab:attack-analysis}
\resizebox{\columnwidth}{!}{%
\begin{tabular}{|p{2.0cm}|p{8.0cm}|}

\hline
\textbf{Attack} & \textbf{Why it Fails} \\
\hline

\textbf{Signature Forgery} & Requires generating $\sigma$ without $\mathsf{sk}$; reduces to EUF-CMA security (Theorem~\ref{thrm:Unforgeability}). \\
\hline

\textbf{Key Extraction} & Violates HBSM guarantees of non-exportability and isolation. \\
\hline

\textbf{Bypassing Liveness} & Requires invoking signing without satisfying policy; prevented by access-controlled interface (Lemma~\ref{lemma:binding}). \\
\hline

\textbf{Sensor Replay / Injection} & Reduces to defeating the liveness mechanism; security depends on the underlying liveness system. \\
\hline

\textbf{Tag Replay} & Duplicate or stale counters are rejected by verifiers (Lemma~\ref{lemma:replay}). \\
\hline

\textbf{Counter Rollback} & Requires violating HBSM-protected state integrity. \\
\hline

\textbf{Server Equivocation} & Detectable via transparency log consistency (Lemmas~\ref{lemma:integrity} \& \ref{lemma:accountability}). \\
\hline

\textbf{Fake Key Registration} & Requires key injection/modification or bypassing attestation verification; prevented by secure registration channels, attestation validation, and trusted server-side log insertion. \\
\hline
\end{tabular}}
\end{center}
\end{table}

\section{Implementation}
\label{sec:implementation}

We implement A-Tag as a cross-platform system with a shared native cryptographic library integrated into Android and iOS applications. The implementation focuses on enforcing hardware-backed key protection and supporting deployment in real-world media-sharing environments.

\textit{Hardware-backed key protection.}
On Android, the system selects the strongest available hardware-backed keystore, prioritizing \emph{StrongBox}~\cite{strongbox}, falling back to the Trusted Execution Environment (TEE)~\cite{armTrustzone}, and rejecting initialization if no hardware-backed support is available. On iOS, A-Tag relies exclusively on the Secure Enclave~\cite{secureEnclave}. This policy ensures private keys remain non-exportable and that signing operations are confined to trusted hardware.

\textit{Deployment considerations.}
Media-sharing platforms strip metadata and apply transformations such as compression, resizing, cropping, and filtering, which break byte-level verification. To preserve verifiability under metadata loss, A-Tag optionally embeds a QR code in the image encoding a reference to the verification endpoint. 
To tolerate content transformations, we optionally incorporate perceptual hashing using PDQ~\cite{pdq}. At tagging time, the system computes three hashes: a full-image hash, an aspect-preserving central crop $(\approx30\%)$, and a fixed $800 \times 800$ central crop. During verification, hashes computed from a candidate image are compared against the stored values, and their aggregate similarity determines whether the image corresponds to the original artifact. This multi-view design improves robustness to common transformations while maintaining discrimination in practice. 
Cryptographic and perceptual hashing serve distinct purposes in A-Tag and are used in separate verification paths. Cryptographic verification establishes the integrity of the exact tagged artifact: any change to its underlying bytes results in a different cryptographic hash and causes verification to fail. In contrast, optional perceptual-hash verification is intended for deployment scenarios in which media-sharing platforms transform images, making byte-level verification impossible. It compares the perceptual characteristics of a candidate image against those of the originally tagged artifact and provides a similarity-based indication of whether the candidate may correspond to that artifact. This comparison is not a cryptographic integrity check and does not establish that the candidate is identical to, or semantically equivalent to, the original image. In particular, localized or adversarial modifications may preserve a high perceptual similarity score. Accordingly, perceptual matching should be interpreted as a tolerance-based identification mechanism rather than cryptographic proof of artifact integrity.

\section{Evaluation}
\label{sec:evaluation}

We evaluate A-Tag on three dimensions, addressing RQ4 and RQ5: (i) retrieval discrimination, (ii) robustness under real-world transformations, and (iii) end-to-end performance on commodity devices.

\subsection{Experimental Setup}

Experiments use 10{,}000 images from the Unsplash Lite dataset~\cite{unsplash}. We distinguish between \emph{offline evaluation} for large-scale retrieval analysis and \emph{on-device execution} for system-level measurements.
For retrieval, we construct \emph{genuine pairs} (images and transformed versions) and \emph{impostor pairs} (different images). Transformations model realistic distortions, including JPEG compression (quality 90–10), resizing (scale 0.90–0.15), central cropping (0.95–0.50 retained), brightness/contrast adjustments (0.50–1.50), and Gaussian blur (radius 0.5–8.0). Parameters are sampled uniformly to generate robustness curves spanning mild to severe perturbations.
In total, we evaluate $\sim$400{,}000 genuine pairs and $\sim$20M impostor pairs. Matching uses Hamming distance between perceptual hashes, evaluated across decision thresholds.
On-device experiments are conducted on an iPhone 14 Pro (iOS 26.5) and a Pixel 6 (Android 16) to measure end-to-end latency and system overhead.

\subsection{Retrieval Accuracy and Robustness}
\begin{figure}[t]
\centering
\includegraphics[width=\linewidth]{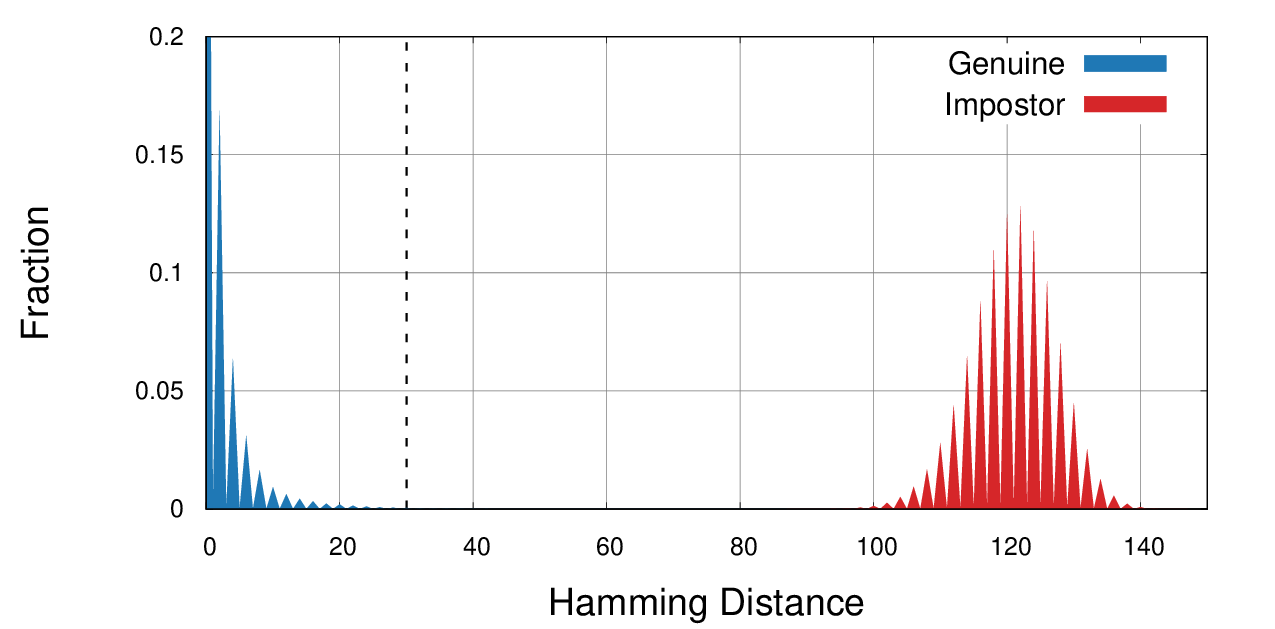}
\caption{Normalized Hamming distance distributions. }
\label{fig:distance-distribution}
\end{figure}

\begin{figure}[t]
\centering
\includegraphics[width=\linewidth]{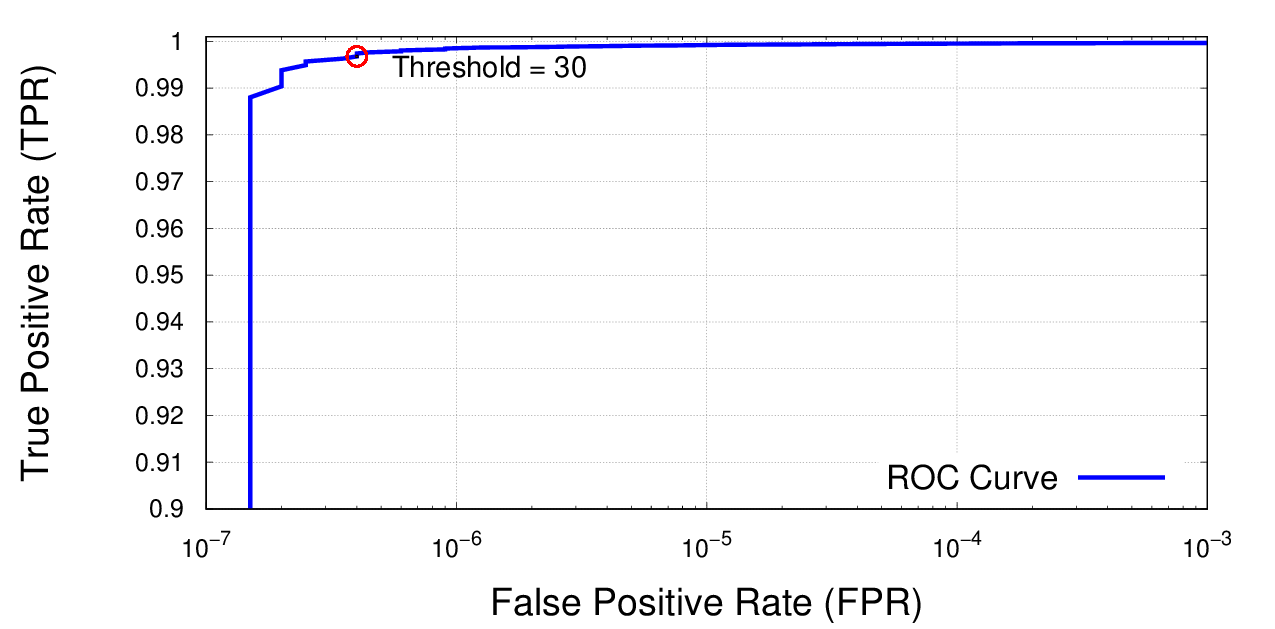}
\caption{ROC curve illustrating the trade-off between true and false positive rates, marking the selected operating point. 
}
\label{fig:roc}
\end{figure}

\begin{figure}[t]
\centering
\includegraphics[width=\linewidth]{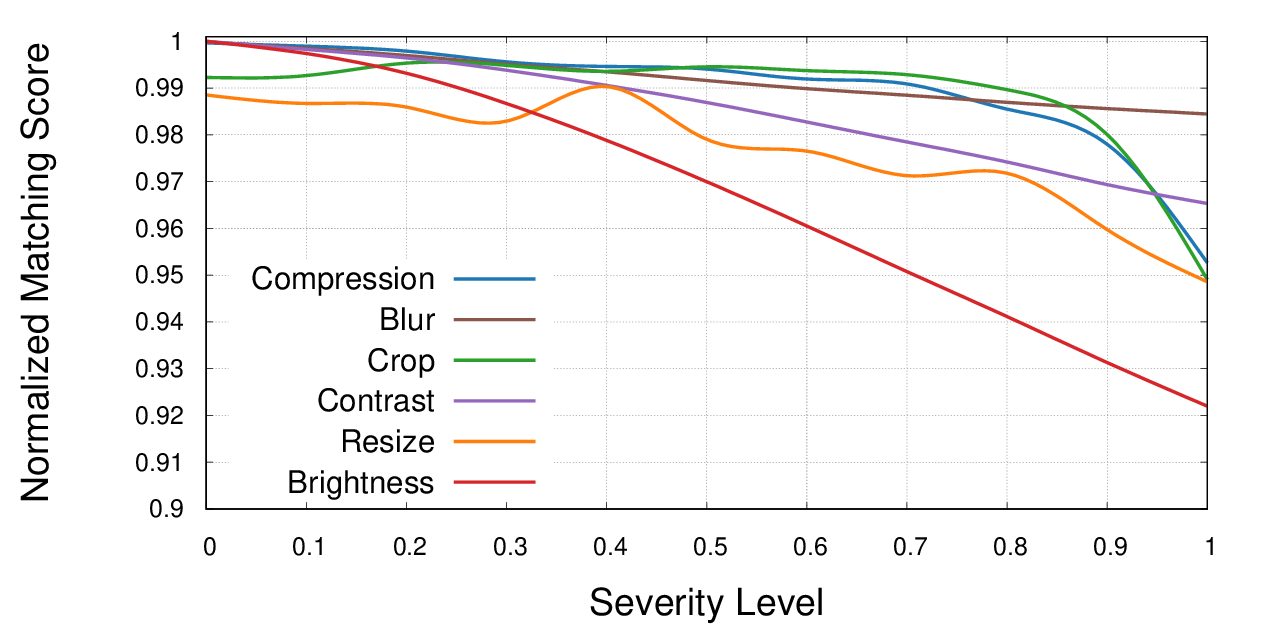}
\caption{Normalized matching score vs.\ transformation severity across families. 
}
\label{fig:robustness-curves}
\end{figure}

\begin{table}[t]
\centering
\caption{Recovery performance for transformation families.}
\label{tab:robustness-summary}
\resizebox{\columnwidth}{!}{%
\begin{tabular}{l c c c c c c}
\toprule
& Compression & Resize & Contrast & Blur & Brightness & Crop \\
\midrule
Recovery (\%) & 100.0 & 100.0 & 100.0 & 100.0 & 99.0 & 99.1 \\
Mean Score ($s$) & 0.988 & 0.963 & 0.975 & 0.988 & 0.950 & 0.988 \\
\bottomrule
\end{tabular}}
\end{table}

We evaluate retrieval performance in terms of discrimination and robustness to image transformations. Figure~\ref{fig:distance-distribution} shows the normalized Hamming distance distributions for genuine and impostor pairs. Genuine pairs concentrate at low distances, while impostor pairs are well separated with minimal overlap. Only 8 out of $\sim$20M impostor pairs fall below 30 bits, all corresponding to duplicate or near-duplicate images present in the dataset. Excluding these, impostor distances begin around 80--100 bits, with the vast majority above 100 bits, yielding a large separation margin. We therefore select a conservative threshold of 30 bits.
Figure~\ref{fig:roc} confirms strong discrimination at this operating point, achieving a TPR of 99.68\% and an FPR of $4.0 \times 10^{-7}$, corresponding to only a handful of false matches across $\sim$20M comparisons.

Robustness is evaluated under common transformations affecting perceptual similarity and variation among genuine pairs. A tag is considered recovered if the Hamming distance falls below the threshold. We report a normalized score $s = 1 - d/n$, where $d$ is the Hamming distance and $n$ is the hash length.
Figure~\ref{fig:robustness-curves} shows the normalized matching score as a function of transformation severity across different distortion families, illustrating graceful degradation under increasing perturbation. Table~\ref{tab:robustness-summary} complements this by reporting recovery rates and mean scores.

Despite strong distortions, recovery exceeds 99\% in most cases, indicating robust performance. Different transformations exhibit distinct degradation profiles. Photometric changes (e.g., brightness) introduce gradual degradation from low severity levels, while geometric distortions such as cropping remain stable initially but degrade more sharply beyond higher severity thresholds due to loss of spatial information. Even under severe distortion, performance remains high; for example, at 50\% of the original image size, recovery decreases to 95.2\%, yet remains robust overall. Excluding such extreme cases, the true positive rate approaches 99\%, indicating that degradation is primarily driven by substantial spatial loss rather than intrinsic limitations of the matching scheme.



\subsection{Latency and System Performance}
We evaluate end-to-end latency by decomposing execution across three components: the A-Tag client, the A-Tag authenticator, and the liveness engine. Table~\ref{tab:latency} summarizes latency across iOS and Android for both base tagging and optional image-based extensions (PDQ computation and QR embedding). Although tags can be verified independently (e.g., by third-party services), we include in-app verification latency for completeness. Internal communication between components is negligible and omitted, while network communication with the A-Tag server is excluded, as it is environment-dependent and orthogonal to intrinsic system performance.

The client introduces negligible overhead for standard inputs and modest overhead for images, where optional PDQ computation and QR embedding dominate. The authenticator is dominated by cryptographic operations, particularly signature generation, and contributes more than the client in the base case but remains small relative to overall end-to-end latency; Android consistently exhibits higher latency than iOS, likely due to differences in hardware and cryptographic acceleration.
The liveness engine dominates tagging latency, with sensor data acquisition accounting for nearly all of its cost, while inference, including feature computation, operates in sub-millisecond time. Consequently, overall latency is driven primarily by external sensing rather than computation. Enabling image-based extensions increases latency due to additional client-side processing but does not change the bottleneck. Across both platforms, end-to-end tagging completes in approximately 1.1--1.4 seconds.
Verification requires only local processing and is dominated by signature verification and matching, with additional overhead when QR decoding and PDQ processing are applied. We report verification latency excluding network communication.

Overall, latency is governed by sensor acquisition rather than cryptographic or hashing operations. As the liveness module is modular and replaceable, this bottleneck is not inherent to A-Tag. Despite incorporating liveness, the system achieves competitive end-to-end latency compared to prior proof-of-personhood approaches that rely on more demanding sensing modalities, while additionally providing cryptographic verifiability and offline validation.

\begin{table}[t]
\centering
\setlength{\tabcolsep}{4pt}

\caption{Latency breakdown by operation (ms).}
\label{tab:latency}
\resizebox{\columnwidth}{!}{%
\begin{tabular}{l l c c c c}
\toprule
\textbf{Component} & \textbf{Operation} 
& \multicolumn{2}{c}{\textbf{iOS}} 
& \multicolumn{2}{c}{\textbf{Android}} \\
\cmidrule(lr){3-4} \cmidrule(lr){5-6}
& 
& \textbf{Base} & \textbf{QR Ext.} 
& \textbf{Base} & \textbf{QR Ext.} \\
\midrule
\multirow{4}{*}{Client}
 & Hash & 0.3 & 0.3 & 0.9 & 0.9 \\
 & PDQ hash & -- & 25 & -- & 67 \\
 & QR embed  & -- & 76 & -- & 79 \\
 & Total latency & 0.3 & 101.3 & 0.9 & 146.9 \\
\midrule
\multirow{2}{*}{Authenticator}
 & Sign & 54 & 54 & 60 & 60 \\
 & Total latency & 123 & 123 & 289 & 289 \\
\midrule
\multirow{3}{*}{Liveness}
 & Sensing & 1000 & 1000 & 1000 & 1000 \\
 & Inference & 0.31 & 0.31 & 0.25 & 0.25 \\
 & Total latency & 1000.31 & 1000.31 & 1000.25 & 1000.25 \\
\midrule
\multirow{1}{*}{Tagging}
 & Total latency & 1123.61 & 1224.61 & 1290.15 & 1436.15 \\
\midrule
\multirow{1}{*}{Verification}
 & Sign verify& 220 & 245 & 420 & 487 \\
\bottomrule
\end{tabular}}
\end{table}

\section{Related Work}
\label{sec:related}
\begin{table*}[!t]
\centering
\caption{Comparison of prior approaches across key properties. 
$\cmark$ = supported, $\sim$ = partial, $\xmark$ = not supported.}
\label{tab:comparison}
\resizebox{\textwidth}{!}{
\begin{tabular}{l l c c c c c c c}
\toprule
\textbf{Category}
& \textbf{System / Approach}
& \textbf{Key}
& \textbf{Origin}
& \textbf{Live}
& \textbf{Action}
& \textbf{Audit}
& \textbf{Human}
& \textbf{Privacy} \\
&
& \textbf{Protection}
& \textbf{Binding}
& \textbf{Presence}
& \textbf{Binding}
&
& \textbf{Evidence}
& \textbf{Preserving} \\
\midrule

\multirow{3}{*}{\textit{Biometric Liveness}}

& PAD Methods~\cite{pad-survey}
& \xmark & \xmark & \cmark & \xmark & \xmark & $\sim$ & \xmark \\
& Face Recognition~\cite{casia-fasd}
& \xmark & \xmark & \cmark & \xmark & \xmark & $\sim$ & \xmark \\
& Behavioral Biometrics~\cite{touch-dynamics}
& \xmark & \xmark & \cmark & \xmark & \xmark & $\sim$ & \xmark \\
\midrule

\multirow{3}{*}{\textit{Proof-of-Personhood}}
& Worldcoin~\cite{worldcoinWhitePaper}
& \xmark & \cmark & \xmark & \xmark & \xmark & $\sim$ & \xmark \\
& Idemix~\cite{idemix}
& \xmark & \cmark & \xmark & \xmark & \xmark & $\sim$ & \cmark \\
& CAPTCHA~\cite{captcha}
& \xmark & \xmark & $\sim$ & \xmark & \xmark & \xmark & \xmark \\
\midrule

\multirow{3}{*}{\textit{Hardware-Backed Auth}}
& ARM TrustZone~\cite{armTrustzone}
& \cmark & \cmark & \xmark & \xmark & \xmark & \xmark & $\sim$ \\
& WebAuthn / FIDO2~\cite{webauthn}
& \cmark & \cmark & $\sim$ & \xmark & \xmark & \xmark & \cmark \\
& Intel SGX~\cite{intelSgx}
& \cmark & \cmark & \xmark & \xmark & \xmark & \xmark & $\sim$ \\
\midrule

\multirow{2}{*}{\textit{Transparency}}
& Certificate Transparency~\cite{laurie2013ct}
& \xmark & \cmark & \xmark & \xmark & \cmark & \xmark & \xmark \\
& CONIKS~\cite{melara2015coniks}
& \xmark & \cmark & \xmark & \xmark & \cmark & \xmark & \xmark \\
\midrule

& \textbf{A-Tag (This Work)}
& \textbf{\cmark}
& \textbf{\cmark}
& \textbf{\cmark}
& \textbf{\cmark}
& \textbf{\cmark}
& \textbf{\cmark}
& \textbf{\cmark} \\

\bottomrule
\end{tabular}
}
\end{table*}

Prior work spans four largely orthogonal directions for establishing trust in digital operations: (i) biometric liveness detection, (ii) proof-of-personhood and Sybil-resistance mechanisms, (iii) hardware-backed key protection, and (iv) transparency frameworks. While some systems partially combine these properties (e.g., liveness-gated authentication~\cite{webauthn, pad-survey} or blockchain-based identity systems~\cite{sybil, worldcoinWhitePaper}), existing approaches provide guarantees only along individual dimensions—liveness detection, uniqueness or human participation, key protection, or auditability—without establishing whether a live human was present at the moment a cryptographic operation was executed.
In particular, liveness is typically verified only at authentication time, after which access is granted and subsequent operations may proceed without further human involvement. This separation enables automated processes or malware to invoke cryptographic operations on behalf of an authenticated user. Moreover, widely deployed authentication systems augmented with liveness detection have been shown to be vulnerable to low-cost physical spoofing attacks~\cite{sp23}. As a result, prior approaches do not provide a verifiable, privacy-preserving, and identity-independent binding between real-time human presence and individual cryptographic operations. A-Tag addresses this limitation by enforcing liveness at the moment of the cryptographic operation and cryptographically binding this evidence to the resulting artifact.

\textbf{Biometric Liveness Detection.} 
Biometric liveness detection distinguishes live human input from spoofing using physiological or behavioral signals, including presentation attack detection (PAD) methods~\cite{livenessSurvey, pad-survey, alive}. While effective in authentication settings, such systems are not cryptographically bound to subsequent operations and cannot guarantee live human presence when a cryptographic action is performed. Furthermore, these systems have been shown vulnerable to increasingly sophisticated spoofing attacks~\cite{sp23}.

\textbf{Proof-of-Personhood and Sybil Resistance.} 
Systems in this space aim to ensure either uniqueness or human participation without relying on real-world identity, often as a foundation for Sybil resistance. Examples include biometric uniqueness systems such as Worldcoin~\cite{worldcoinWhitePaper}, anonymous credential systems~\cite{idemix}, and challenge-response mechanisms for transient human participation such as CAPTCHA~\cite{captcha}. While these approaches establish either uniqueness or transient human participation, they do not bind such guarantees to individual cryptographic operations.

\textbf{Hardware-Backed Authentication.} 
Authentication frameworks such as WebAuthn/FIDO2~\cite{webauthn} rely on hardware-backed authenticators for key protection and origin binding guarantees. These authenticators typically use secure hardware, including HBSMs and TEEs such as ARM TrustZone~\cite{armTrustzone} and Intel SGX~\cite{intelSgx}. However, they only capture coarse-grained user presence signals (e.g., simple interaction checks) and do not produce verifiable evidence of human participation during cryptographic operations. Additionally, prior work demonstrated vulnerabilities in complex TEEs, particularly in system-level components with large attack surfaces~\cite{foreshadow}. This motivates designs that minimize reliance on complex environments and instead employ narrowly scoped hardware-backed key protection with reduced functionality and attack surface.

\textbf{Transparency Frameworks.} 
Transparency systems such as Certificate Transparency and CONIKS~\cite{laurie2013ct, melara2015coniks} ensure public auditability of key lifecycle events. While they provide strong guarantees of accountability and detect misbehavior in key management, they remain agnostic to how cryptographic keys are used in practice and do not capture human involvement.

Table~\ref{tab:comparison} summarizes these approaches across key security properties, including key protection, origin binding, live presence, action binding, auditability, human evidence, and privacy preservation. We distinguish between detecting live human presence at interaction time (\emph{Live Presence}) and producing cryptographically verifiable evidence of such participation (\emph{Human Evidence}). The comparison shows that prior systems provide strong guarantees along individual dimensions; however, none establish a verifiable, privacy-preserving binding between real-time human presence and individual cryptographic operations. In contrast, A-Tag uniquely satisfies all properties simultaneously.

\section{Conclusion}
\label{sec:conclusion}

We presented A-Tag, a system binding digital artifacts to cryptographic signing operations performed with a live human present. By combining hardware-backed key protection, liveness-gated signing, and transparency-backed registration, A-Tag guarantees unforgeability, human presence, and public verifiability, shifting key-management trust from opaque intermediaries to publicly auditable transparency logs. Our implementation and evaluation show these guarantees are achievable efficiently on commodity devices with minimal overhead. Future work includes strengthening robustness against advanced sensor-level attacks and extending optional confidence-based image verification to videos undergoing platform-level compression and transformations.

\bibliographystyle{ACM-Reference-Format}
\bibliography{References}

\end{document}